\documentclass[11pt,twoside]{./atmp}

\usepackage{amsmath,amssymb}

\newcommand{\rmd}{{\rm d}}
\newcommand{\rme}{{\rm e}}
\newcommand{\rmi}{{\rm i}}

\copyrightnotice{2007}{11}{0}{24} 

\begin{document}

\title[A mathematical framework for a standard theory]
{ A mathematical framework \\for a standard theory \\using extended representations \\of paths and world lines }

\arxurl{gr-qc/0702022}

\author[B.\ H.~Dunford-Shore]{B.\ H.~Dunford-Shore}

\address{~12430 Tesson Ferry Road \#203, Saint Louis, Missouri 63128-2702}  
\addressemail{STANDPHYS@earthlink.net}

\begin{abstract}
An analysis using a composition of currently-accepted theories is given. 
Starting with a synthesis of what may be generically termed ``paths'', analysis
of representations for these ``paths'' is developed.  Foreground and background
interactions are explicitly treated by using a local representation
that treats the two representations equally and symmetrically.  A restriction
coupling from the global space-time representation to local interaction source
terms is treated in terms of mass and charge couplings. Rewriting the connection
in terms of the global manifold and the coupled terms yields compatibility with
Dirac and Klein-Gordan equations for electro-weak coupled particles and fields.
Compatibility with currently-accepted theories that includes standard charge
assignments, SU(3) confinement, and a definition for particle flavor
generations is used to constrain and validate the composition and the analysis.
\end{abstract}

\maketitle

\cutpage 

\setcounter{page}{2}

\noindent

\section{%
 \label{sec:intro}
 Introduction
}

This paper will try to address the following question: to what extent is it
possible to take the elements of currently-accepted theories that are
experimentally verified and combine them to produce real, consistent physics? 
This assumes proceeding with the minimum possible usage of speculative higher
symmetries and groups, dimensions, new particles, and fields.  The idea is to
combine existing elements in ways compatible with quantum field theory and
general relativity to the extent that this is possible.

Some key practices
include: working with known symmetries and groups such as U(1), SU(2), SU(3),
and SO(3), working with four or 3+1 dimensions so that a cross product is
possible, working with a holomorphic, analytic-continuation representation,
working with
classical
concepts such as phase and configuration space trajectories (paths) and world
lines, and not assuming any new particle or field content.  It must be
acknowledged that, currently, any composition of the key elements of accepted
theories is, in a real sense, speculative in its own right.

\subsection{%
 \label{ssec:introsynthesis}
 Synthesis
}

It is already known that phase trajectories (also known as phase paths) and
world lines are concepts that are compatible with each other\cite{Goldstein}.  
The paths (or orbits or trajectories) of particles in space-time, their phase or
configuration space paths, and world lines are different representations or
aspects of the same thing.  These can also be understood using the concepts of
symmetry group (gauge) transformations and orbits.  Another type of fundamental
representation that is essential is field representations such as probability
wave function representations.  These and their associated machinery such as
Lagrangians and Hamiltonians are all complementary methods that express and
manage different aspects or facets of what is being studied.

Some other aspects that are important to consider are whether the
representations are local or global, and whether the underlying representation
(metric) is considered to be flat or curved.  Analytic-continuation
representations and Wick
rotations also provide means of representing and working with different aspects.
 The most realistic approach is to combine the usage of different methods and
representations in ways that allow the different aspects to be handled best.

To have the best chance of fitting different elements of current theories
together, a generalized analysis will be given.  The local representations will
be setup to maintain consistency with current theories.  The relationships
required by current theories will then be imposed.  This allows for the best
consistency with current theories while allowing for any implicit relationships
between current theories to become manifest.

Compatibility with the currently-accepted theories will provide constraint and
validation for the development of the synthesis and analysis.  The compatibility
of phase paths with quantum field theory, when used carefully, will be used as
the basis for maintaining compatibility with both quantum field theory and
general relativity.  Analytic-continuation representations and Ashtekar's new
variables
\cite{PhysRevLett.57.2244} will also be assumed to help form a common basis
between quantum field theory and general relativity.  In the interest of
brevity, the statement of the ties to currently-accepted theories will be
restricted to the minimum necessary to constrain the analysis and validate
compatibility.

\subsection{%
 \label{ssec:introapproach}
 Path, representation, and combination approach
}

The approach this paper will use will be to treat paths, orbits, trajectories,
and world-lines of the various sorts such as phase, configuration, space-time,
field, gauge group, or relativistic as similar entities whose representations
capture the differences in the facets or aspects of the different entities.  To
this end, a synthesis that abstracts the treatment of ``paths'' and their
``representations'' will be done.  The probability relations (functions and
functionals) for these entities are also an important aspect or facet.

This approach starts by considering numbers to be capable of being decomposed
into two ``dual'' parts--each part being tangent and cotangent four-dimensional
representations.  The next step is to consider merging these numbers together to
create extended objects--but preferring that the numbers be put together only
if their representations merge together smoothly and continuously.  These 
(complex) numbers are identifiable as system points and their extended objects
as paths.  The local representations are constructed as symmetric,
holomorphic, analytic-continuation
representations so that either representation can be a foreground or background
representation.  This allows full foreground and background (self) interaction
to be present.  The smooth merging of the paths and their local chart
representations allows these local charts to be ``glued'' together to form a
global manifold solution\cite{WaldGR}.  The global manifold solution is a
background representation that does not participate in the local foreground and
background symmetry.  The global manifold solution is given as a restriction of
the holomorphic, analytic-continuation (symmetric) representation.

Permutations of the possible paths, the possible representations for the paths,
and combinations and transitions of the representations constitute an initial
local superset realm\cite{smolin1990lsr}.  Not all of this realm can be
effectively realized as (directly) observable solutions. We will follow the
convention of calling the observable entities ``effective'' entities. Other
entities we will call either ``initial'' or ``local'' entities. We assume
initial entities have only local structure while effective entities have both
local and global structure.

For full foreground and background symmetry, holomorphic, analytic-continuation
representations for the
paths are introduced.  Explicit mappings between the conjugate representations
is given.  Local representation quantities are expressed conjugate to each other
at each point but the physical Hilbert structure is only imposed on the
effective representation\cite{smolin1990lsr} and is not considered in this
paper.  Local interactions within and between these representations are
explored.  These representations are then mapped (restricted) to a target
space-time manifold solution. 

When the combinatorial sum composition is taken, some global solutions have the
possibility of separating so that they have no direct coupling to the local
interactions.  Most other global solutions do not have the local interaction
coupling eliminated.  The coupled remainder for the local interactions leads to
what may be identified as mass terms. The particles have local interaction terms
that couple to the mass unless they are capable of the cancellations necessary
to decouple and to become independent of any mass terms (or other local coupled
terms).

\subsection{%
 \label{ssec:conventions}
 Conventions
}

We will not follow the convention of setting $\hbar = 1$, $c = 1$, $m_e = 1$,
etc.  It is also part of the purpose of this paper to track the origin and
combinations of the fundamental constants.  That is easier to do when they are
explicit in the equations.  The range (dimensionality) of indexes will be
specified in the paper when it is appropriate to assume a restricted range,
except that the indices $i$ and $j$ are assumed to run from one to three.

\section{%
 \label{sec:paths}
 Paths (and trajectories and world lines)
}

An essential ingredient in the mathematics for physics is the combination of
entities that produce a (possibly complex) number or numeric (scalar) field as
a result.  The general form in mathematics is to map two entities to a number. 
The map of vectors $V$ and differential forms $\omega$ to (possibly complex)
numbers $Z$, $V : \omega \rightarrow Z$, will be the basic form considered here.
 The remainder of this section will
state some general machinery for considering ``paths''.

To analyze paths, start with an unordered set of complex numbers that does not
contain zero: $Z \subset \mathbb{C}$, $0 \not\in Z$.  Please do not yet think of
these numbers $Z$ as points as there is not yet an associated space or manifold.
Consider different permutated subsets of the values in the set $Z$ of $\lambda
\subset Z$, in which $\lambda \neq \emptyset$.  Define these (range) subsets as
the images of different functions $\lambda(\omega)$.  Define the functional that
is  constructed by using a domain set consisting of all functions whose (range)
images $\lambda$ are all the possible  permutations of (domain) subsets of $Z$
\begin{equation}
 \label{eq:Lambda}
 \Lambda[\Omega] = \{\lambda(\omega)~|~\\
 \forall z, \forall \lambda, z = \lambda(\omega), \\
 z \in \lambda\}.
\end{equation}
The $\lambda(\omega)$ are then defined as path functions whose values
are the complex numbers $z$ in some permutation.  There is not \textit{a priori}
a requirement for the values along a path $\lambda(\omega)$ to be smooth or
continuous--in the general case, they could jump around in any order and be
discontinuous.

The path element $\triangle\lambda$ is a discrete numerical representation.  The
differential form $\rmd\lambda$ represents an infinitesimal (continuous) piece
of
numeric path.  For ease of notation, represent both $\triangle\lambda$ and
$\rmd\lambda$ by $\rmd\lambda$
and create a path
\begin{equation}
 \label{eq:path}
 \lambda_n(\omega) = \sum_m \rmd\lambda_{n(m)}(\rmd\omega_{n(m)})
\end{equation}
in which $n(m)$ is some different arrangement of the $\lambda$.  As defined so
far, these paths are possibly very chaotic--allowing continuous and discrete
changes along the path and different and incompatible representations of
elements along the path.  In the general case, the different
representation spaces for each path element could include changes in scale,
orientation, dimension, and structure such as algebra and groups.

\section{%
 \label{sec:representations}
 Representations
}

Giving quantum field theory preeminent place, the probability representation for
a path will be considered to be a primary representation.  The problem of
foreground and background representation interaction will next be explicitly
addressed. Differential form representations are considered next in importance
with their particular representations constrained or validated by compatibility
with current theory.

\subsection{%
 \label{ssec:probwavefunction}
 Probability amplitude functions
}

For probability representations it is important to define probability amplitude
functions and functionals.  Define $\Phi[\Lambda]$ as a functional whose domain
is the
set of functions that assigns complex probability amplitude images (the
function's range)
to each path
$\lambda$ (the function's domain):
\begin{equation}
 \label{eq:Phi}
 \Phi[\Lambda] = \{\phi(\lambda)~|~\forall c, c = \phi(\lambda), c \in
\mathbb{C}\}.
\end{equation}
Allow the value of the probability amplitude to be everywhere zero for
particular paths $\lambda$.  In other words, some particular paths will be so
pathological that they do not effectively exist.

\subsection{%
 \label{ssec:foreback}
 Foreground and background representations
}

A key problem to address explicitly up-front is the problem of foreground and
background definitions and their interactions.  The goal is to try to treat the
foreground and background representations equally and symmetrically. We will
start as usual with local definitions called coordinate patches or
charts\cite{WaldGR}.  Define $V$ as the foreground component.  Define
$\rmd\omega$
in differential form as the background representation that has the conjugate
coordinate basis $\frac{\partial{}}{\partial{\omega}}$.  These definitions are
such that $V : \rmd\omega \rightarrow \lambda$ and  \begin{equation}
 \Phi^\dagger(\lambda) \frac{\partial{\Phi(\lambda)}}{\partial{\omega}} = V,
\end{equation}
where $\Phi(\lambda)$ is the probability amplitude wave function of the entity
that is to be represented.

Now consider allowing a portion $B$ of the foreground $V = g_B B V_1$ to act on
the
background representation
\begin{equation}
 \Phi^\dagger(\lambda) B \frac{\partial{\Phi(\lambda)}}{\partial{\omega}} = V_2.
\end{equation}
Write $B$ in differential form $B = b(\rmd\omega_B) \rmd\omega_B$ and write a
combined
holomorphic, analytic-continuation representation using the conjugate of
$\rmd\omega_B$
\begin{equation}
 \Phi^\dagger(\lambda) ( \frac{\partial{}}{\partial{\omega}} + \rmi g_B
\vec{\tau_B}
\frac{\partial{}}{\partial{\omega_B}} )\Phi(\lambda) = V_2.
\end{equation}
The use of a dimensional constant $g_B$, placing, the
$\frac{\partial{}}{\partial{\omega_B}}$ representation on the imaginary axis,
and/or usage of a group vector $\vec{\tau_B}$ can be important to make sure that
there is not inappropriate mixing of terms.  If it is appropriate, then $g_B$
and
$\vec{\tau_B}$ can be set to one.

Another way of viewing this is that the ``foreground'' definition is a
representation in its own right and to treat it
symmetrically\cite{nasiri2006psq}.  To this end, two
local representations, $\rmd\omega_1$ and $\rmd\omega_2$, will be introduced in
which
one
can be thought of as a local foreground definition and the other as a local
background definition--but with nothing (yet) selecting which one is which. 
This type of formulation allows us to try to simultaneously fit both the
$\rmd\omega_1$ and $\rmd\omega_2$ representations to smooth transitions.  In the
case
that $\omega_1$ is a spatial coordinate, then included in $\omega_2$ are
conjugate momentum terms that correspond to translational transformations of
$\omega_1$ and conjugate angular momentum terms that correspond to rotational
transformations of $\omega_1$.  This program of extending the representations
corresponds to a program of embedding the possible transformations of a
representation into a larger, extended representation.  The
internal and external transformations will also be embedded into extended
representations with (gauge) fields providing a mapping restriction back to a
restricted representation.

Write a holomorphic, analytic-continuation representation in terms of coordinate
patch bases
\begin{subequations}
\begin{equation}
 \label{eq:diffbasisomega}
 \frac{\partial{}}{\partial{\omega}} = \frac{\partial{}}{\partial{\omega_1}} +
\rmi
g_\omega \vec{\tau_\omega} \frac{\partial{}}{\partial{\omega_2}}.
\end{equation}
or
 \begin{equation}
 \label{eq:diffbasis}
 \partial = \partial_1 + \rmi g_\omega \vec{\tau_\omega} \partial_2.
\end{equation}
\end{subequations}

Quantum conditions can be assumed by excluding a region of the complex plane
(which could be located at the origin).  If an area of a particular size
$\rmd\omega_1
\rmd\omega_2$ is excluded from the complex plane, then the quantum condition
\begin{equation}
 \label{eq:heisenberg}
 \rmd\omega_1^a \rmd\omega_2^a \geq \hbar / 2
\end{equation}
has been applied (no summation on $a$--these are separate conditions).  For
classical conditions, either take the limit as $\hbar$ goes to zero or scale up
to a scale very much larger than $\hbar$ so that $\hbar$ 1is effectively
negligible (zero).

To highlight the key differences from the usual definitions, start by
considering a standard definition of a bundle $\mathcal{B}$ with manifold
$\mathcal{M}$, fiber $\mathcal{V}$, and group $\mathcal{G}$.  Next we see that
for local, full symmetry we want a ``bundle'' with fibers $\mathcal{V}_1$ and
$\mathcal{V}_2$--not the usual definition of a bundle.  In the case that
foreground and background symmetry is ignored, then $\mathcal{V}_1$ may be
mapped to a (global) manifold $\mathcal{M}$ and a standard bundle $\mathcal{B}$
results.

\subsection{%
 \label{ssec:exttransform}
 Transformations on holomorphic representations
}

This subsection states some properties of transformations and gauge field
representations of holomorphic representations\cite{thiemann1996atg}. 
Please
consider external (group) transformations $\Theta_\mathcal{E} \in \mathcal{G_E}$
operating on the representation
\begin{equation}
 \label{eq:Domega}
 D_\omega = \Theta_\mathcal{E} \partial = \Theta_\mathcal{E} ( \partial_1 + \rmi
g_\omega \vec{\tau_\omega} \partial_2 ).
\end{equation}
These correspond to the internal (conjugate) transformations that transform
between $\omega_1$ and $\omega_2$.  These transformations are such that
$\Theta_\mathcal{E} : \omega_1 \rightarrow \omega_2$ or $\Theta_\mathcal{E} :
\omega_2 \rightarrow \omega_1$ which, when summed (if
possible) are
\begin{subequations}
 \begin{equation}
 \label{eq:fourier1}
 f'(\omega_2) = \int \Theta_\mathcal{E} f(\omega_1) \rmd\omega_1
\end{equation}
and
 \begin{equation}
 \label{eq:fourier2}
 f(\omega_1) = \int \Theta_\mathcal{E} f'(\omega_2) \rmd\omega_2.
\end{equation}
\end{subequations}
If $g_\omega = 1$, $\vec{\tau_\omega} = 1$, and the $\Theta_\mathcal{E}$
transformations are U(1) external transformations $\rme^{\pm \rmi \theta}$ that
correspond to the internal (conjugate) transformations $\frac{1}{\sqrt{2 \pi}}
\rme^{\pm \rmi \omega_2 \omega_1}$ then (\ref{eq:fourier1}) and
(\ref{eq:fourier2}) are Fourier and inverse Fourier transformations as
(internal) rotations in the complex plane.  If the group structure of
$\vec{\tau_\omega}$ corresponds to some group SU(N), of which SU(2) or SU(3) are
used in this paper, then corresponding (internal) transformation equations for
(\ref{eq:fourier1}) and (\ref{eq:fourier2}) may be obtained.

Defining the ``momentum'' (gauge field) representation that is conjugate to the
external transformations $\Theta_\mathcal{E}$ representation as $\vec{B}$ gives
another form of representation
\begin{equation}
 \label{eq:fieldrep}
 D_B = \partial + \rmi g \vec{\tau} \vec{B}.
\end{equation}

\subsection{%
 \label{ssec:localglobalrep}
 Local and global representations couplings
}

The preceding representations are local definitions.  These are partial
derivatives with no transition functions (yet) defined to other (local)
derivative representations (coordinate patches or charts).  For a global
definition, a preferred background representation will have to be identified and
(summed combination) transition functions and operations will have to be
identified and used.

The global representation is a (restriction) subset of the degrees of freedom
and interactions that can be present locally.  When going from local to global
representations, some local (internal) symmetries or group transformations may
show closure--the group orbits within their phase or configuration volumes may
be shrunk to ``points'' from a global perspective.  The local interactions
become hidden and some part of the system acts as a composite single entity. 
This means that not all quantities that are present locally are quantities that
must be representable directly on the global representation--especially with
full foreground and background interaction symmetry. These local representations
can
be written as a composite representation term $\epsilon$ that couples to the
global space-time representation.  This coupling of a global manifold to locally
coupled terms can be written as
\begin{equation}
 \label{eq:backgroundtolocal}
 \Phi^{*} {D^\mathcal{L}}_a \Phi \leftrightarrow
 \Psi^{*} D_{X^a} \Psi + \epsilon(\lambda)
\end{equation}
where $\epsilon(\lambda)$ are the local
coupled terms and $D_{X^a} = \frac{\partial{}}{\partial{X^a}}$ are the
coordinates of a global manifold.

\section{%
 \label{sec:local}
 Local representations
}

For two sets of entities $\omega_1$ and $\omega_2$ that have maps to numbers
$\Upsilon : \Omega \rightarrow Z$, define two complementary conjugate
representations.  Define a mapping on $V_1$ to $\rmd\lambda \in
\rmd\lambda_{n(m)}$,
\begin{equation}
 \label{eq:ZPmap}
 \rmd\omega_1 : V_1 \rightarrow \rmd\lambda,
\end{equation}
so that $\rmd\omega_1 V_1 = \rmd\lambda$, in which $V_1 \in \Upsilon$ and $V_1
\not\in
\Omega$.  Also define a mapping from $V_2$ to a $\rmd\lambda$
\begin{equation}
 \label{eq:ZXimap}
 V_2 : \rmd\omega_2 \rightarrow \rmd\lambda
\end{equation}
so that $V_2 \rmd\omega_2 = \rmd\lambda$, in which $V_2 \in \Omega$ and $V_2
\not\in
\Upsilon$.  The terms $V_1$ and $V_2$ are assumed to be complex vectors and the
terms $\rmd\omega_1$ and $\rmd\omega_2$ are assumed to be complex differential
forms. 
They are assumed to have properties local to $\rmd\lambda_{n(m)}$.

\subsection{%
 \label{ssec:dformrep}
 Differential form representation
}

For each $\rmd\omega$ there are many possible choices for their representation
as
coordinate one-form representations
\begin{equation}
 \label{eq:omegaxi}
 \rmd\omega = u(\xi^a) \rmd\xi^a
\end{equation}
with the numeric tuples $u(\xi^a)$.  In the general case, these different
choices of coordinates
could include differences in scale, orientation, dimension, and structure such
as algebra and groups depending on different choices for $\xi$.

Some relations that follow are:
\begin{subequations}
 \begin{equation}
 \label{eq:dlambdaVdxi}
 \rmd\lambda = V \rmd\omega = V_a u(\xi^a) \rmd\xi^a,
\end{equation}
which can be integrated to give
\begin{equation}
 \label{eq:lambdaVdxi}
 \lambda(\xi) = \int V_a u(\xi^a) \rmd\xi^a,
\end{equation}
and whose derivative with respect to $\xi^a$ can be taken to give
\begin{equation}
 \label{eq:dlambdadxi}
 \frac{\partial{\lambda(\xi)}}{\partial{\xi^a}} = V_a u(\xi^a).
\end{equation}
Using (\ref{eq:Phi}) gives
\begin{equation}
 \label{eq:dpsidxi1}
 \phi^\dagger(\lambda) \frac{\partial{\phi(\lambda)}}{\partial{\xi^a}} =
\phi^\dagger(\lambda) \frac{\rmd\phi(\lambda)}{\rmd\lambda} V_a u(\xi^a),
\end{equation}
\end{subequations}
in which $\frac{\rmd\phi(\lambda)}{\rmd\lambda}$ is a total derivative since
$\phi$ is
only a function of $\lambda$.  In general, the $\phi(\lambda)$ might be
chosen to be in a form in which $\frac{\rmd\phi(\lambda)}{\rmd\lambda}$
contributes
factors such as $\hbar$ and $\pm \rmi$.  Unless stated otherwise, in this paper
it
will be assumed that $\phi(\lambda)$ are in a form in which
\begin{equation}
 \label{eq:phionlylambda}
 \phi^\dagger(\lambda) \frac{\rmd\phi(\lambda)}{\rmd\lambda} = 1
\end{equation}
and
\begin{equation}
 \label{eq:dpsidxi}
 \phi^\dagger(\lambda) D_{\xi^a} \phi(\lambda) =
 \phi^\dagger(\lambda) \frac{\partial{\phi(\lambda)}}{\partial{\xi^a}} = V_a
u(\xi^a).
\end{equation}

\subsection{%
 \label{ssec:repassumptions}
 Compatible representation assumptions
}

For ease of notation, the representation of $\omega_1$ will be given as $\chi^a$
and the representation of $\omega_2$ will be given as $\pi^a$ and if the
discussion applies to both equally, their representations will be referred to as
$\xi^a$.  The conjugate for $\chi^a$ will be given as $\kappa_a$ and the
conjugate for $\pi^a$ will be given as $\Xi^a$.

For compatibility, assume four dimensions for $\chi^a$, $\pi^a$, and $\xi^a$ and
all related dimensional quantities for the remainder of the paper.  It is fairly
standard to treat the four dimensions as 3+1 dimensions.  Use the zero index for
the separate dimension and give $a$ and $b$ the ranges
\begin{equation}
 \label{eq:muD}
 a, b = 0, 1, 2, 3.
\end{equation}

For $\xi^a$ (for each of $\chi^a$ and $\pi^a$), arbitrarily pick one axis
$\xi^0$
and assume a (scaled)
identity alignment mapping within the different $\xi^0$'s.  Assume the
functional derivatives between possible coordinate basis sets is
\begin{equation}
 \label{eq:sigmaxi}
 g_\mu \sigma^\mu = \frac{\delta{\xi^a}'}{\delta{\xi^a}}
\end{equation}
(no summation), in which $\sigma^0 = \textbf{1}$, the $\sigma^i$ are Pauli
matrices, and the $\sigma^\mu$ are a representation of SU(2). Assume Gell-Mann
matrices $\lambda_1$, $\lambda_2$, and $\lambda_3$ that are isomorphic to the
matrices $\sigma^i$ and that belong to the group $G_{s}(3)$ of SU(3) rotations
on the $\xi^i$ axes.

Since the zero axis was arbitrarily picked, there is a gauge transformation in
SU(2) for the $\xi$ basis that corresponds to multiplying through by an element
of SU(2) to pick a different $\xi^0$ axis associated with the identity element. 
Designate by $\vec{\tau_\mu}$ the set of SU(2) elements for this gauge
transformation.

The $\xi^a$ ($\chi^a$ and $\pi^a$) bases are local to each point $z$ with
partial derivatives that commute and, therefore, locally flat metrics $\eta_{a
b}$.  The possible metric signatures for $\xi^a$ are assumed to be $[-,+,+,+]$
or $[+,-,-,-]$ ($\xi^a$ is indexed from zero).  These correspond (respectively)
to the $\xi^i$ ($\chi^i$ and $\pi^i$) being positive definite and having a
right-handed chirality or negative definite and having a left-handed chirality.

\subsection{%
 \label{ssec:groupstructmapxi}
 Group closure restriction to idempotent basis
}

Define the mapping $G_{s} : \xi^i \rightarrow \xi^0$ so that
\begin{equation}
 \label{eq:upsilonxi}
 \sum_i \frac{\partial{\xi^i}}{\partial{\xi^0}} = \pm G_s \vec{\upsilon_\xi}.
\end{equation}
The positive or negative sign comes from the metric signature and $G_s$ is a new
dimensional constant that ensures $\vec{\upsilon_\xi}$ is a vector value.

Using the rotation $\lambda_n$, identity combinations of the trivial
representation can be constructed.  The group trivial representation
$\lambda_0 = \textbf{1}$ is
\begin{equation}
 \label{eq:Dixi}
 {}^{I}\!D_{\xi^0} = \frac{\partial{}}{\partial{\xi^0}}.
\end{equation}
The group inverse representation $\lambda_n^{-1} \lambda_n = \textbf{1}$ gives
\begin{equation}
 \label{eq:Dgixi}
 {}^{GI}\!D_{\xi^0} = G_s^{-1} \vec{\upsilon_\xi}^{~-1}
 G_s \vec{\upsilon_\xi} \frac{\partial{}}{\partial{\xi^0}}.
\end{equation}
The group closure representation $\lambda_i \lambda_j \lambda_k = \textbf{1}$
gives
\begin{equation}
 \label{eq:Dgcxi}
 {}^{GC}\!D_{\xi^0} = -G_s \vec{\upsilon_\xi}^{~i}
 G_s \vec{\upsilon_\xi}^{~j}
 G_s \vec{\upsilon_\xi}^{~k}
 \frac{\partial{}}{\partial{\xi^0}},
\end{equation}
in which $i \neq j$, $j \neq k$, $i \neq k$.  The representation on $\xi^a$
that does not export a dependency on $\lambda_n$ (an interaction with
$\lambda_n$) is the one whose group element is $\lambda_0$.  A subset
representation of $\xi^a$ that does not have an exported dependency on
$\lambda_{a\neq 0}$ is the combination of the group identity as given by
(\ref{eq:Dixi}), the group inverse as given by (\ref{eq:Dgixi}), and the group
closure as given (\ref{eq:Dgcxi}).  This is
\begin{equation}
 \label{eq:Dxi}
 D_{\xi^0} = ~{}^{I}\!D_{\xi^0} + ~{}^{GI}\!D_{\xi^0} + ~{}^{GC}\!D_{\xi^0}.
\end{equation}

\subsection{%
 \label{ssec:extrepresentations}
 Transformation based extended representations
}

As given in section \ref{ssec:exttransform}, transformations $\Theta_\mu$
between
the two representations $\rmd\chi$ and $\rmd\pi$ can be written as additional
representations
\begin{equation}
 \label{eq:thetaB}
 \Theta_\mu \partial \rightarrow D_B = \partial + \rmi g_B \vec{\tau_\mu}
\vec{B_a^\mu}.
\end{equation}
Including the U(1) identity orientation $\vec{\tau_0} = \textbf{1}$, there are
four possible orientations $\vec{\tau_\mu}$ between $\rmd\chi$ and $\rmd\pi$. 
Corresponding to the four $\vec{\tau_\mu}$ there are different maps $\Theta_\mu
: \rmd\chi \leftrightarrow \rmd\pi$, $\Theta_\mu = \{ \vartheta_\mu, \zeta_\mu
\}$
between $\xi_{\chi}^0$ and $\xi_{\pi}^0$:
\begin{subequations}
\begin{equation}
 \label{eq:gvartheta}
 \vec{\tau_\mu} \frac{\partial{\xi_{\pi}^0}}{\partial{\xi_{\chi}^0}}
 = g_\vartheta \vartheta_\mu
\end{equation}
and
\begin{equation}
 \label{eq:gvartheta2}
 \vec{\varsigma_\mu}
 \frac{\partial{\xi_{\chi}^0}}{\partial{\xi_{\pi}^0}} = g_\zeta \zeta_\mu,
\end{equation}
\end{subequations}
with corresponding vector potentials $\vec{B_a^\mu}$ and $\vec{{B'}_a^\mu}$.

The translational transformations and rotational transformations for the target
manifold coordinates will be assumed to already be included in the momentum
conjugate to the target manifold coordinates.  The SU(2) basis for $\omega_{1
i}$, has SU(3) transformations so that 
\begin{equation}
 \label{eq:flavor}
 G_f \partial_{\omega_1 i} \rightarrow D_F = \partial_{\omega_1 i} + \rmi g_F
\vec{\tau_F} \vec{B_F}.
\end{equation}
The SU(2) basis for $\omega_{2 i}$, also has SU(3) transformations so that 
\begin{equation}
 \label{eq:strong}
 G_S \partial_{\omega_2 i} \rightarrow D_S = \partial_{\omega_2 i} + \rmi g_S
\vec{\tau_S} \vec{B_S}.
\end{equation}

To contain the scope of the paper, unless stated otherwise in the remainder of
the paper, the restricted forms of (\ref{eq:flavor}) and (\ref{eq:strong})
in the form given by (\ref{eq:Dxi}) will be used.

\subsection{%
 \label{ssec:mapped}
 Mapped representations
}

A mapped representation can be written in terms of transformation fields
$\Theta_\mu$ as
\begin{subequations}
\begin{equation}
 \label{eq:localderivative}
 {D^\mathcal{A}}_a  = \Theta_\mu
 [ (g_{\zeta} \zeta_\mu)^m \vec{\varsigma^\mu} D_{\chi^a}
 + (g_\vartheta \vartheta_\mu)^n \vec{\tau^\mu} D_{\pi^a} ],
\end{equation}
in which $m \in \{ 0, 1 \}$, $n \in \{ 0, 1 \}$, $n \neq m$ and
\begin{equation}
 \label{eq:Theta}
 \Theta_\mu = \prod \varpi_\mu, ~\varpi_\mu \in \{ \vartheta_\mu, ~\zeta_\mu \}.
\end{equation}
\end{subequations}
This expresses one of the representations as a native representation
and the other as a mapped representation.  To preserve foreground and background
symmetry by not writing a preferred form, write an analytic extension
\begin{equation}
 \label{eq:fghlocal}
 {D^\mathcal{L}}_a = {D^\mathcal{A}}_a - \rmi {D^\mathcal{B}}_a
\end{equation}
by the use of (\ref{eq:localderivative}) and
\begin{equation}
 \label{eq:DLprime}
 {D^\mathcal{B}}_a = 
 {\Theta'}_\mu
 [ (g_{\zeta} \zeta_\mu)^{m'} \vec{\varsigma^\mu} D_{\chi^a}
 + (g_\vartheta \vartheta_\mu)^{n'`} \vec{\tau^\mu} D_{\pi^a} ],
\end{equation}
in which it is assumed that the $m \neq m'$ and $n \neq n'$.

Equations (\ref{eq:Theta}) and (\ref{eq:fghlocal}) allow transformations
(rotations) between representations ${D^\mathcal{A}}_a$ and ${D^\mathcal{B}}_a$
of ${D^\mathcal{L}}_a$ by taking the proper transformations $\Theta_\mu$ and
${\Theta'}_\mu$ (which include the U(1) transformations of the form $\rme^{\rmi
\theta}$) to give a transformed representation
\begin{equation}
 \label{eq:dualrotation}
 {D^\mathcal{L'}}_a = \delta\Theta {D^\mathcal{L}}_a,
\end{equation}
in which $\delta\Theta$ is a functional transformation of $\Theta_\mu$ and
${\Theta'}_\mu$.

\section{%
 \label{sec:localterms}
 Local coupled terms
}

Up to this point, local, full symmetry has been considered with a ``bundle''
with fibers $\mathcal{V}_1$ and $\mathcal{V}_2$.  At this point, consider
mapping $\mathcal{V}_1$ to a (global) manifold $\mathcal{M}$ and a standard
bundle $\mathcal{B}$ with manifold
$\mathcal{M}$, fiber $\mathcal{V}$, and group $\mathcal{G}$.

If the phase and configuration spaces $\alpha_{R}$ for this restriction to the
bundle $\mathcal{B}$ are separable from the extended possibilities of the local,
full symmetry $\alpha_{R}$, then the properties of the bundle are intrinsic.  If
the phase and configuration spaces $\alpha_{R}$ for this restriction to the
bundle $\mathcal{B}$ are not cleanly separable from the extended possibilities
(phase and configuration space $\alpha_{F}$) of the local, full symmetry, then
the properties of the bundle will be considered to have remaining extrinsic
terms $\epsilon$.  This section will define terms $\epsilon$ with the intrinsic
and extrinsic possibilities built into the definition.

Define the unit configuration space volume element as the volume element in
which each dimension has a unit value of $\hbar c$ for a volume element of
$\alpha_{unit} = (\hbar c)^N$.

\subsection{%
 \label{ssec:localenergy}
 Local internal interaction coupling
}
Designate ``local internal interaction'' (real) terms as
\begin{eqnarray}
 \label{eq:localmomentum}
   \epsilon_m(\lambda) & = & E_m M(\lambda) \left(\begin{array}{c}
    L_M(\alpha_{\mathtt{I}}) \\
    L_M(\alpha_{\mathtt{E}})
  \end{array}\right) \\
 \nonumber
 & = & E_m M(\lambda) \iota_M,
\end{eqnarray}
in which $\alpha_{\mathtt{I}}$ and $\alpha_{\mathtt{E}}$ are intrinsic and
extrinsic configuration volume values and $E_m$ contains a dimensional coupling
constant $m$ to these terms.  Assume $m$ is scaled equal to  the electron mass
$m = m_e$. Assuming a velocity of $c$ and scaling over a unit value of
$\hbar c$, lets us assume the term is $E_m = \frac{m c^2}{\hbar c}$. Assume the
boundary of the configuration volume for a particular $D_{x^a}$ for intrinsic
$L_M(\alpha_{\mathtt{I}})$ is such that the local coupled configuration volume
is zero.  In this case, the $D_{x^a}$ decouples from the local configuration
volume so that $\alpha_{\mathtt{I}} = 0$ and $L_M(\alpha_{\mathtt{I}}) = 0$. 
Otherwise, it is the case that $\alpha_{\mathtt{E}} \neq 0$ and
$L_M(\alpha_{\mathtt{E}}) = 1$.  The ``local internal interaction'' coupling
term is then written as 
\begin{eqnarray}
 \label{eq:localmomentummc}
   \epsilon_m(\lambda) & = & \frac{m c^2}{\hbar c} M(\lambda)
\left(\begin{array}{r}
    0\\
    1
  \end{array}\right) \\
 \nonumber
 & = & \frac{m c^2}{\hbar c} M(\lambda) \iota_M.
\end{eqnarray}

\subsection{%
 \label{ssec:localalignment}
 Local external interaction coupling
}

Any global target manifold need not be aligned with a particular
${D^\mathcal{L}}_a$ corresponding to $\Theta_\mu$.  Designate the local
difference in alignment as a ``local external interaction'' alignment variance
\begin{eqnarray}
 \label{eq:localalignment1}
   \epsilon_e(\lambda) & = & \Theta_e Q(\lambda) \left(\begin{array}{c}
    L_Q(\Theta_\mu)\\
    L_Q(\Theta_\mu)
  \end{array}\right) \\
 \nonumber
 & = & \Theta_e Q(\lambda) \iota_Q,
\end{eqnarray}
in which $\Theta_e$ contains a dimensional coupling constant that is normalized
over $\hbar c$ of
\begin{equation}
 \label{eq:ecoupling}
 \Theta_e = \pm \frac{e}{\hbar c}.
\end{equation}
Assume $L_Q(\Theta_\mu) = 1$.  The representation alignment coupling term is
then
\begin{eqnarray}
 \label{eq:localalignment}
   \epsilon_e(\lambda) & = & \pm \frac{e}{\hbar c} Q(\lambda)
\left(\begin{array}{r}
    1\\
    1
  \end{array}\right) \\
 \nonumber
 & = & \pm \frac{e}{\hbar c} Q(\lambda) \iota_Q.
\end{eqnarray}
The $\Theta_\mu$ transformation is properly both positive and negative because
it rotates between chirality--it rotates between metric signatures of
$[-,+,+,+]$ for $D_{\chi^a}$ and $[+,-,-,-]$ for $D_{\pi^a}$.

\subsection{%
 \label{ssec:localremainder}
 Total local interaction coupling
}

The total local coupled ``local interaction'' terms for
$\Phi^{*}
{D^\mathcal{L}}_a \Phi$ is
\begin{equation}
 \label{eq:localsenergy}
   \epsilon(\lambda) = \epsilon_m(\lambda) {\Theta^\mathcal{M}}_\mu
 - \rmi \epsilon_e(\lambda) {\Theta^\mathcal{E}}_\mu.
\end{equation}
The $\epsilon(\lambda)$ are coupled to $\Phi$ by way of ${D^\mathcal{L}}_a$
but local fields do not have to go through this coupling to interact with other
local fields.

\section{%
 \label{sec:global}
 Global manifold
}

Introduce entities $x^A$, $p_A$, $X^a$, $P_a^{A}$, and ${W}_a^\mu$ where $X^a$
and $P_a^{A}$ are assumed to have diffeomorphic, smooth, continuous transition
functions on the intersections of the $x^A$ and $p_A$.  The $x^A$ and $p_A$ are,
respectively, tangent space and cotangent space effective representations of the
smooth, continuous, effective phase path solutions $\lambda(X,P) \in Z$.  The
$X^a$ are the coordinates of the global manifold solution and are a transform
solution from the $x^A$, which are variants of the $\chi^a$.  The ${W}_a^\mu$
are
the equivalent transformation from the $B_a^\mu$ that map to $x^A$ instead of to
$\chi^a$ and $\pi^a$. Equations equivalent to (\ref{eq:ZPmap}) and
(\ref{eq:omegaxi}) for $X^a$ are a mapping from $X$ to $Z$ of
\begin{subequations}
\begin{equation}
 \label{eq:Xkappamap}
 X : P \rightarrow Z
\end{equation}
so that $X P = Z$ with a representation of
\begin{equation}
 \label{eq:x}
 X = \{X^a~|~X^a = g(X^a) \rmd X^a, g(X^a) \in \mathbb{C} \}
\end{equation}
\end{subequations}
and the numeric tuple $g(X^a)$.

Assume there is an isomorphism $\sigma_{\chi} \rightarrow \sigma_X$ and that the
$i \sigma_X$ are the generators of the group SO(3) of rotations on the $X^i$
axes.  We will use the standard convention for writing the $x$ coordinate
representation
\begin{equation}
 \label{eq:Dx}
 D_{x^a} = \gamma^a \partial_a = \gamma^a \frac{\partial}{\partial{x^a}},
\end{equation}
with the manifold coordinate representation written as
\begin{equation}
 \label{eq:DX}
 D_{X^a} = \gamma^a \partial_{X a} = \gamma^a \frac{\partial}{\partial{X^a}},
\end{equation}
in which $\gamma = \varsigma \otimes \sigma_X$ are the Dirac matrices and
$\varsigma$ is a factor from (\ref{eq:dualrotation}).

Equations equivalent to (\ref{eq:dlambdaVdxi}),(\ref{eq:lambdaVdxi}), and
(\ref{eq:dpsidxi}) for
$X^a$ and $P$ are (where $\mathbf{e}^A$ are $x^A$ orthonormal frame sets)
\begin{subequations}
\begin{equation}
 \label{eq:dlambdakappadx}
 \mathbf{e}^A \rmd\lambda = P_a^{A} \rmd X^a,
\end{equation}
\begin{equation}
 \label{eq:lambdakappadx}
 \mathbf{e}^A \lambda(X,P) = \int P_a^{A} g(X^a) \rmd X^a,
\end{equation}
and
\begin{equation}
 \label{eq:dpsidx}
 \mathbf{e}^A \psi^\dagger(X,P) \frac{\partial{\psi(X,P)}}{\partial{X^a}} =
g(X^a) P_a^{A}.
\end{equation}
\end{subequations}

Writing the distance along an effective phase path as a world line as
\begin{subequations}
 \begin{equation}
 \label{eq:distance}
 \rmd s^2 = \rmd\lambda^{*} \rmd\lambda =
 g^{*}(X^a) g(X^b) {P^{*}}_a^{A} P_b^{B} \eta_{A B}
 {\rmd X^{*}}^a \rmd X^b
\end{equation}
gives
 \begin{equation}
 \label{eq:distancemetric}
 \rmd s^2 = g_{a b} {\rmd X^{*}}^a \rmd X^b,
\end{equation}
in which the metric has the form\cite{PhysRevLett.57.2244}
 \begin{equation}
 \label{eq:metric}
 g_{a b} = g^{*}(X^a) g(X^b) {P^{*}}_a^{A} P_b^{B} \eta_{A B}
\end{equation}
\end{subequations}
and $P_a^A$ is in the form of an Ashtekar tetrad ``soldering form'' that links
to the neigh1borhood of tangent spaces $x_A$ that define the manifold value of
$X_a$.  Because a neighborhood $X^a$ of the manifold may be composed of
multiple
coordinate patch sets of ($x^a$,~$p_a$), $P_a^A$ can represent a value that is
not on the mass-shell.

\subsection{%
 \label{ssec:localglobal}
 Composite coupling
}

The restriction to a background manifold derivative fixes the direction of
${W}_a^\mu$ and transforms $\Xi_a$ and $\kappa_a$ to the coupled ``local
interaction''
\begin{eqnarray}
 \label{eq:mapbackground}
 \Phi^{*} {D^\mathcal{L}}_a \Phi & \rightarrow &
 \Psi^{*} D_{x^a} \Psi + \Phi^{*} ( {D^\mathcal{A}}_a - \rmi {D^\mathcal{B}}_a)
\Phi
\\
 \nonumber
 & = &
 \Psi^{*} D_{x^a} \Psi + \epsilon(\lambda),
\end{eqnarray}
so that (using (\ref{eq:localsenergy}))
\begin{equation}
 \label{eq:omegaDL}
 \Phi^{*} ({D^\mathcal{A}}_a - \rmi {D^\mathcal{B}}_a) \Phi =
 \epsilon_m(\lambda) {\Theta^\mathcal{M}}_\nu - \rmi
\epsilon_e(\lambda) {\Theta^\mathcal{E}}_\nu.
\end{equation}

\subsection{%
 \label{ssec:localinternal}
 Internal interaction coupling
}

Using (\ref{eq:localderivative}), the real part transforms (restricts) as a
variation of $\chi^a$ to give the manifold and a term $\epsilon_m(\lambda) =
\Psi^{*} {D^\mathcal{A}}_a \Psi$ in which
\begin{equation}
 \label{eq:backgroundchi1}
 \epsilon_m(\lambda) {\Theta^\mathcal{M}}_\nu = 
 \Phi^{*}  ( \varsigma_{\nu_0} D_{\chi^a}
 + g_\vartheta \vartheta_{\nu_1} \vec{\tau^{\nu_1}} D_{\pi^a} ) \Phi
\end{equation}
so that $m = 0$, $n = 1$.  An alternate transform is
\begin{equation}
 \label{eq:background}
 \epsilon_m(\lambda) {\Theta^\mathcal{M}}_\nu = 
 \Phi^{*} [ g_\vartheta \vartheta_{\nu_1} (
 g_{\zeta} \zeta^{\nu_0} \vec{\varsigma^{\nu_0}} D_{\chi^a} 
 + \vec{\tau^{\nu_1}} D_{\pi^a} ) ] \Phi
\end{equation}
so that $m = 1$, $n = 0$.  Fix the gauge of the corresponding gauge field $Z_a$
by assuming a gauge fixing transformation $\vartheta_{\nu_1} \rightarrow 0$,
${\Theta^\mathcal{M}}_\nu = 1$.  Using (\ref{eq:localmomentummc}) gives
\begin{equation}
 \label{eq:mprime}
 \epsilon_m(\lambda) = \Phi^{*} \varsigma_{\nu_0} D_{\chi^a} \Phi =
 \frac{m c^2}{\hbar c} M(\lambda) \iota_M.
\end{equation}

\subsection{%
 \label{ssec:localexternal}
 External interaction coupling
}

Using (\ref{eq:DLprime}), the imaginary part transforms (restricts) to give
\begin{equation}
 \label{eq:backgroundxi}
 \epsilon_e(\lambda) {\Theta^\mathcal{E}}_\nu = 
 \Phi^{*} [ g_\vartheta \vartheta_{\nu_3} (
 g_{\zeta} \zeta_{\nu_2} \vec{\varsigma^{\nu_2}} D_{\chi^a}
 + \vec{\tau^{\nu_3}} D_{\pi^a} ) ] \Phi
\end{equation}
so that $m' = 1$, $n' = 0$.  Rewriting using (\ref{eq:localalignment}) gives
\begin{equation}
 \label{eq:qprime}
 \frac{e}{\hbar c} Q(\lambda) \iota_Q =
 \Phi^{*} [ g_\vartheta ( g_{\zeta} \zeta_{\nu_2} \vec{\varsigma^{\nu_2}}
D_{\chi^a}
 + \vec{\tau^{\nu_3}} D_{\pi^a} ) ] \Phi,
\end{equation}
in which the gauge field $A_a$ is introduced as
\begin{equation}
 \label{eq:thetae}
  {\Theta^\mathcal{E}}_\nu = \vartheta_{\nu_3} = \gamma^a A_a.
\end{equation}

\section{%
 \label{sec:compatibility}
 Standard Model compatibility
}

This section will establish compatibility with the Standard Model in terms of
electroweak fields, charges, and Dirac and Klein-Gordon equations.

\subsection{%
 \label{ssec:maptransform}
 Interaction field
}
Arbitrarily assigning $W^{-}_a$ the index 1 location and $W^{+}_a$ the index
2 location, the new ${W}_a^\mu$ may be given as
\begin{subequations}
\begin{equation}
 \label{eq:W}
 {W}_a^\mu = ( A_a, W^{-}_a, W^{+}_a, Z_a ),
\end{equation}
in which the Standard Model
form\cite{RevModPhys.52.515,RevModPhys.52.525,RevModPhys.52.539} is
assumed:
\begin{equation}
 \label{eq:A}
 A_a = \sin \theta_W B^3_a + \cos \theta_W B^0_a,
\end{equation}
\begin{equation}
 \label{eq:Wboson}
 W^\pm_a = \frac{B^1_a \pm \rmi B^2_a}{\sqrt{2}},
\end{equation}
and
\begin{equation}
 \label{eq:Z}
 Z_a = \cos \theta_W B^3_a - \sin \theta_W B^0_a.
\end{equation}
\end{subequations}
It is reasonable to assume $A_a$ has the form
\begin{equation}
 \label{eq:Awork0}
   A_a \left(\begin{array}{rr}
    1 & 0\\
    0 & 0
  \end{array}\right)
\end{equation}
so that, using (\ref{eq:localmomentummc}), it is massless
\begin{equation}
 \label{eq:Amass}
 A_a \left(\begin{array}{rr}
    1 & 0\\
    0 & 0
  \end{array}\right) \epsilon_m(\lambda) = 
 A_a \left(\begin{array}{rr}
    1 & 0\\
    0 & 0
  \end{array}\right) 
   \frac{m c^2}{\hbar c} M(\lambda) \left(\begin{array}{r}
    0\\
    1
  \end{array}\right) = 0.
\end{equation}

Charge is a perpendicular transformation (rotation) in the direction of the
local axis ${D^\mathcal{B}}_a$, with $\epsilon_e(\lambda)$ and $A_a$ having
perpendicular rotation in the direction of that axis, and with $A_a$ centered
(gauge fixed) at zero.  The $A_a$ has zero net charge and has a rotation
alignment of zero with respect to the $x$ manifold.  The $A_a$ is
four-dimensional and has no mass coupling from (\ref{eq:Amass}).  Mass is
along the local real axis ${D^\mathcal{A}}_a$, with $\epsilon_m(\lambda)$ and
$Z_a$ along that axis, and with $Z_a$ displaced from zero.  The $Z_a$ has no
charge coupling and has a mass. The $W^\pm_a$ are along, and off-axis, for both
axes.  The $W^\pm_a$ have a charge coupling and have masses.  The $W_a^{+}$, its
antiparticle $W_a^{-}$, and $Z_a$ can be easily shown to couple to mass using
(\ref{eq:localmomentummc}).

It is possible that the preceding description does not preserve complete Lorentz
invariance, but SIM(2) plus strong CP preservation (local Lorentz invariance) as
given in Very Special Relativity\cite{cohen2006vsr} should be preserved.

The assignments that implement what has just been discussed are to assign $g_A =
g_{\vartheta}$ and set $\zeta^{\nu_0} = \zeta^0$, $\vec{\varsigma^{\nu_0}} =
\vec{\varsigma^{0}}$, $\zeta^{\nu_2} = \zeta^3$, $\vec{\varsigma^{\nu_2}} =
\vec{\varsigma^{3}}$, and $\vec{\tau^{\nu_3}} = \vec{\tau^{0}}$.  Rewriting
(\ref{eq:mprime}) with these assignments gives
\begin{equation}
 \label{eq:mprimef}
 \epsilon_m(\lambda) = \Phi^{*} \vec{\varsigma^{0}} D_{\chi^a} \Phi =
 \frac{m c^2}{\hbar c} M(\lambda) \iota_M.
\end{equation}
Rewriting (\ref{eq:qprime}) with these assignments gives
\begin{equation}
 \label{eq:qprimeref}
 \frac{e}{\hbar c} Q(\lambda) \iota_Q =
 \Phi^{*} ( g_A g_{\zeta} \zeta^3 \vec{\varsigma^{3}} D_{\chi^a}
 + g_A \vec{\tau^{0}} D_{\pi^a} ) \Phi.
\end{equation}

The new ``direction'' of ${W}_a^\mu$ is set relative to $x^0$.
Since $D_{x^a}$ is a variation (restriction transformation) of $D_{\chi^a}$,
assume
\begin{equation}
 \label{eq:gA}
 \Phi^{*} g_A \vec{\tau^{0}} \Phi = \frac{1}{2} \left(\begin{array}{r}
    1 \\
    1
  \end{array}\right).
\end{equation}

\subsection{%
 \label{ssec:weakmixingangle}
 Weak mixing angle
}

The (restriction) transform from $B_a^\mu$ to ${W}_a^\mu$ goes from groups
SU(2) to SU(2)/U(1) with a preferred duality rotation direction $e$. The charge
$e$ is a scale factor with respect to $\theta_W$--the $g'/g$ set the direction
of $\theta_W$ and $e$ is the (inverse unit) magnitude along the direction.  The
charge $e$ then effects a scaling to make it a partial (restriction)
transformation instead of a complete (restriction) transformation.  The charge
is then as given in the Standard Model as $e^2 = g^2 \sin^2 {\theta}_{W}$ with
directions $\sin^2 {\theta}_{W} = e^2/g^2$, $\cos^2 {\theta}_{W} = e^2/{g'}^2$,
and $\tan^2 {\theta}_{W} = g'^2/g^2$ in which $\sin^2 \theta_W$ is the Standard
Model\cite{RevModPhys.52.515,RevModPhys.52.525,RevModPhys.52.539} weak mixing
angle.

\subsection{%
 \label{ssec:transformationassignments}
 Charge assignments
}

Using (\ref{eq:qprimeref}) and (\ref{eq:gA}) lets us write 
\begin{equation}
 \label{eq:chargebase0}
   \Phi^{*} g_A \vec{\tau^{0}} D_{\pi^0} \Phi
 = -\frac{1}{2} \frac{e}{\hbar c} \left(\begin{array}{r}
    1 \\
    1
  \end{array}\right)
 = \frac{e}{\hbar c} \left(\begin{array}{r}
    -\frac{1}{2} \\
    -\frac{1}{2}
  \end{array}\right),
\end{equation}
in which there is a change of sign for $\Phi^{*} D_{\pi^0} \Phi$ due to the sign
of the metric of $\pi$.  Writing the term for $D_{\pi^i}$ with a factor of
$\frac{1}{3}$ to account for the three terms of $\pi^i$ gives
\begin{equation}
 \label{eq:chargebasei}
   \Phi^{*} g_A \vec{\tau^{0}} D_{\pi^i} \Phi 
 = \frac{1}{3} \frac{1}{2} \frac{e}{\hbar c} \left(\begin{array}{r}
    1 \\
    1
  \end{array}\right)
 = \frac{e}{\hbar c} \left(\begin{array}{r}
    \frac{1}{6} \\
    \frac{1}{6}
  \end{array}\right).
\end{equation}
Using (\ref{eq:qprimeref}) allows us to write 
\begin{equation}
 \label{eq:chargefunctionleft}
   \Phi^{*} g_{A} g_{\zeta} \zeta^3 \vec{\varsigma^{3}} D_{\chi^a} \Phi 
 = \frac{1}{2} \frac{e}{\hbar c} 
 \left(\begin{array}{r}
    1 \\
    -1
  \end{array}\right)
 = \frac{e}{\hbar c} \left(\begin{array}{r}
    \frac{1}{2} \\
    -\frac{1}{2}
  \end{array}\right).
\end{equation}

Using (\ref{eq:qprimeref}) and (\ref{eq:chargefunctionleft}) and using
(\ref{eq:chargebase0}) for leptons and using (\ref{eq:chargebasei}) for
quarks then gives Standard Model compatible charge assignments:
\begin{itemize}
\item A charge for leptons
\begin{equation}
 \label{eq:Qlepton}
 Q(\lambda_{\Xi^0}) \iota_Q =
  \left(\begin{array}{r}
    -\frac{1}{2} \\
    -\frac{1}{2}
  \end{array}\right)
 + \left(\begin{array}{r}
    \frac{1}{2} \\
    -\frac{1}{2}
  \end{array}\right)
 = \left(\begin{array}{r}
    0 \\
    -1
  \end{array}\right).
\end{equation}
\item A charge for quarks
\begin{equation}
 \label{eq:Qquark}
 Q(\lambda_{\Xi^i}) \iota_Q =
 \left(\begin{array}{r}
    \frac{1}{6} \\
    \frac{1}{6}
  \end{array}\right)
 + \left(\begin{array}{r}
    \frac{1}{2} \\
    -\frac{1}{2}
  \end{array}\right)
 = \left(\begin{array}{r}
    \frac{2}{3} \\
    -\frac{1}{3}
  \end{array}\right).
\end{equation}
\end{itemize}

\subsection{%
 \label{ssec:connection}
 Dirac and Klein-Gordon equations
}

The restriction transformation (\ref{eq:mapbackground}) may be rewritten as
\begin{equation}
 \label{eq:mapbackgroundfunction}
 \Phi^{*} {D^\mathcal{L}}_a \Phi
  \rightarrow 
 \Psi^{*} D_a \Psi,
\end{equation}
in which
\begin{equation}
 \label{eq:DG1}
 D_a =  D_{x^a}
 + \frac{m c^2}{\hbar c} M(\lambda) \iota_M
 - \rmi \gamma^a A_a \frac{e}{\hbar c} Q(\lambda) \iota_Q.
\end{equation}
Substituting (\ref{eq:Qlepton}) into (\ref{eq:DG1}) gives, for leptons,
\begin{equation}
 \label{eq:DGXI0}
 {}_{\Xi_0}\!D_{a} =
 \gamma^a \partial_a
 + \frac{m c^2}{\hbar c} M(\lambda) \left(\begin{array}{r}
    0\\
    1
  \end{array}\right)
 - \rmi \gamma^a A_a \frac{e}{\hbar c} \left(\begin{array}{r}
    0 \\
    -1
  \end{array}\right).
\end{equation}
Splitting (\ref{eq:DGXI0}) into its two parts gives a neutrino equation,
\begin{equation}
 \label{eq:Dneutrino}
 \Psi^{*} {D^\mathcal{V}}_a \Psi = \Psi^{*} \gamma^a \partial_a \Psi,
\end{equation}
and an electron-like equation (in which ${D^\mathcal{E}}_a$ stands for all three
``electron-like'' particles),
\begin{equation}
 \label{eq:Delectron}
 \Psi^{*} {D^\mathcal{E}}_a \Psi =
 \Psi^{*} ( \gamma^a \partial_a
 + \frac{m c^2}{\hbar c} M(\lambda) 
 + \rmi \gamma^a A_a \frac{e}{\hbar c} ) \Psi
\end{equation}
This shows that the three neutrinos can decouple and the electron, muon, and
tauon do not.  The coupled local interactions $\epsilon(\lambda)$ are seen to
contain all of the source terms.

The equation for the global derivative is (a type of Dirac
equation\cite{DiracPMQM}), from (\ref{eq:DG1}),
\begin{equation}
 \label{eq:dirac}
 \Psi^{*} D_a \Psi = \Psi^{*} [ \gamma^a (\partial_a - \frac{i
e}{\hbar
c}
 q(\lambda_\Xi) A_a )
 + \frac{m c^2}{\hbar c} m(\lambda_\kappa) ] \Psi,
\end{equation}
in which $q(\lambda_\Xi)$ are allowed combinations of $Q(\lambda_{\Xi^0})$ and
$Q(\lambda_{\Xi^i})$ and $m(\lambda_\kappa)$ are allowed combinations of
$M(\lambda)$. When dealing with the group inverse terms of $\pi$ (scalar
equations), a Klein-Gordan form is more appropriate:
\begin{equation}
 \label{eq:kleingordan}
 \Psi^{*} D_a^{*} D_a \Psi = \Psi^{*}
 (\partial_a^2 + \frac{e^2}{\hbar^2 c^2} q(\lambda_\Xi)^2 A_a^2 
 + \frac{m^2 c^4}{\hbar^2 c^2} m(\lambda_\kappa)^2 ) \Psi.
\end{equation}

\section{%
 \label{sec:totals}
 Totals and time
}

So far in this paper, care has been taken to not identify any particular
dimension or variable as time.  The 3+1 dimensions and group structures that
have been used so far are compatible with the selection of a preferred time
dimension but full symmetry has been maintained.  Some sort of parameterization
of the paths in terms of an explicit separate parameter or in terms of time
could be assumed to allow usage of the functional path
integral\cite{RevModPhys.20.367,seidewitz2007fsp}.  This is done to make sure
that the integrals are well defined and do not diverge.

For the path representation overlap to be smooth, the paths must be properly
ordered and joined together in sets of compatible ``histories''. Also, for the
overlap to be smooth, the transition functions for path representations must be
smooth or differentiable.  Another way to put this is that the path
representations must be diffeomorphic at the overlap.  If the path
representations are (indirectly) diffeomorphic to a common representation then
that establishes a diffeomorphic relationship between them.  The diffeomorphism
of the common representation to a total derivative such as time helps to
parameterize and order the permutations of the paths.  The presence of a total
derivative such as time implies smoothness and continuity of representations and
paths that select for permutations of paths that are effective.  Therefore, the
paths that cannot be parameterized in terms of a monotonically increasing
parameter must be excluded from the set of considered paths.

\subsection{%
 \label{ssec:composites}
 Composite entities
}

The configuration volume can contain products of terms such as $\lambda_{\chi a}
\kappa^a$ and $\lambda_{\pi a} \Xi^a$.  Constructing enclosed configuration
subspaces for these give the terms defined in \eqref{eq:Dgixi} and
\eqref{eq:Dgcxi}. It is possible to assume $\lambda_{\chi a}$ and $\lambda_{\pi
a}$ are confined symmetries by assuming that the external transformations
$\Theta_\mu$ do not project $\lambda_{\chi a} \kappa^a$ and $\lambda_{\pi a}
\Xi^a$ onto the underlying basis--that these terms disappear in the restriction
projection
\begin{equation}
 \label{eq:confinementchi}
 \Theta_\mu \kappa^a = 0,~a \neq 0
\end{equation}
and
\begin{equation}
 \label{eq:confinementxi}
 \Theta_\mu \Xi^a = 0,~a \neq 0.
\end{equation}
The content of this assumption is that the external transformations $\Theta_\mu$
interact
only with the zero-indexed terms $\kappa^0$ and $\Xi^0$.

The following dimensional basis-state constructs are then possible:
\begin{itemize}
 \item A four-dimensional configuration consisting of only
($x^a$,~\nolinebreak$p^a$).  The local configuration volume coupling
$\epsilon(\lambda)$ is zero.
 \item A five-dimensional configuration consisting of the ($x^a$,
~\nolinebreak$p^a$) and also $\epsilon(\lambda)$ coupled  to the
($\Xi^0$,~\nolinebreak$\kappa^0$) terms.
 \item A seven-dimensional configuration consisting of the
($x^a$,~\nolinebreak$p^a$) and also $\epsilon(\lambda)$ coupled to the
($\Xi^i$,~\nolinebreak$\kappa^i$) terms.  Combinations of the
($\Xi^i$,~\nolinebreak $\kappa^i$) terms are first mapped to the
($\Xi^0$,~\nolinebreak$\kappa^0$) terms by (combinations of) group inverse and
group closure restrictions.
 \item Combinations of the terms ($\Xi^a$,~\nolinebreak $\kappa^a$) that do not
map or couple to the ($x^a$,~\nolinebreak $p^a$) and are therefore hidden,
local transition terms.
\end{itemize}

If the $\epsilon(\lambda)$ are zero, then time and energy are intrinsic and the
representation is four-dimensional.  If $\epsilon(\lambda)$ is not zero then,
then effectively the representation has more than four dimensions.  The $\Xi^0$
and $\kappa^0$ constitute part of a five-dimensional representation.  The
$\Xi^i$ and $\kappa^i$ and their group inverses and group closure restrictions
constitute a seven-dimensional representation mapped to five dimensions and then
coupled to four dimensions.

The terms ${}^{I}\!D_{\chi^0}$, ${}^{GI}\!D_{\chi^0}$, and ${}^{GC}\!D_{\chi^0}$
are compatible with characterizing different flavor or generation terms.  The
${}^{I}\!D_{\chi^0}$ term is assumed to correspond to the first generation
particles (electron, electron neutrino, and down and up quarks).  The
${}^{GI}\!D_{\chi^0}$ term is assumed to correspond to the second generation
particles (muon, muon neutrino, and strange and charm quarks).  The
${}^{GC}\!D_{\chi^0}$ term is assumed to correspond to the third generation
particles (tauon, tauon neutrino, and top and bottom quarks).

The terms ${}^{I}\!D_{\pi^0}$, ${}^{GI}\!D_{\pi^0}$, and ${}^{GC}\!D_{\pi^0}$
are compatible with leptons, mesons, and baryons respectively.

\subsection{%
 \label{ssec:energyequation}
 Energy
}

Assume in the standard way that there is a representation (restriction) where
the coordinate $X^0$ is diffeomorphic (with a Wick rotation) to the total
derivative--that the coordinate $X^0$ is diffeomorphic to time
$\frac{\partial{X^0}}{\partial{t}} = \rmi c$. Assume its dual $P^0$ is equal to
the
total energy 
\begin{equation}
 \label{eq:energy}
 P^0 = \rmi E/c.
\end{equation}

Using \eqref{eq:kleingordan},\eqref{eq:dpsidx}, \eqref{eq:metric}, and
\eqref{eq:energy}, assuming the gauge condition $A_a = 0$, and setting
\eqref{eq:kleingordan} equal to a constant
\begin{equation}
 \label{eq:kleingordanvalue}
 \frac{1}{\hbar^2 c^2} \Lambda_C = \Psi^{*} D_a^{*} D_a \Psi
\end{equation}
gives
\begin{eqnarray}
 \label{eq:energy1}
 \frac{1}{\hbar^2 c^2} \Lambda_C & = & 
 g(X_a)^2 P^a P_a + \frac{m^2 c^4}{\hbar^2 c^2} m(\lambda)^2 \\
 \nonumber
 & = & - \frac{g(X_0)^2}{c^2} E^2 + g(X_i)^2 P^i P_i
 + \frac{m^2 c^4}{\hbar^2 c^2} m(\lambda)^2.
\end{eqnarray}
If $g(X_a) = \frac{1}{\hbar}$, then
\begin{equation}
 \label{eq:zeroenergy}
 \frac{1}{\hbar^2 c^2} \Lambda_C = -E^2 \frac{1}{\hbar^2 c^2}
 + P^i P_i \frac{c^2}{\hbar^2 c^2}
 + \frac{m^2 c^4}{\hbar^2 c^2} m(\lambda)^2
\end{equation}
or
\begin{equation}
 \label{eq:totalenergy}
 E^2 + \Lambda_C = P^i P_i c^2 + m^2 c^4 m(\lambda)^2.
\end{equation}
For the neutrino, this becomes $E^2= P^i P_i c^2 - \Lambda_C$.  For the
electron, this becomes $E^2= P^i P_i c^2 + m_e^2 c^4 - \Lambda_C$.  Using a Wick
rotation $\Lambda_C = \rmi \epsilon_F$, gives the standard form needed for the
convergence
of the path integral\cite{nambu1950upt}.

\section{%
 \label{sec:discussion}
 Discussion
}

This paper has demonstrated a holomorphic representation that is
compatible with
general relativity and Dirac and Klein-Gordon equations.  An interpretation
that is compatible with Standard Model group structure, fields, charge
assignments, and particle content has also been shown.

Time is introduced on the physical restriction (effective) representation but
full four-dimensional covariance and group structures are maintained on all of
the ``underlying'' (local and analytic continuation) representations.  The local
foreground and background conjugate symmetry, when restricted to the effective
(physical) representation is seen to be able to impose the proper properties and
structure on the spacetime manifold and its conjugate particle content.

\bibliographystyle{my-h-elsevier}

\begin{thebibliography}{10}

\bibitem{Goldstein}
H. Goldstein,
Classical mechanics,
2d ed. (Addison-Wesley Pub. Co., Reading, Mass., 1980).

\bibitem{PhysRevLett.57.2244}
A. Ashtekar,
New Variables for Classical and Quantum Gravity,
Phys. Rev. Lett. {\bf 57} (1986) 2244.

\bibitem{WaldGR}
R.M. Wald,
General relativity,
(University of Chicago Press, Chicago, 1984).

\bibitem{smolin1990lsr}
L. Smolin and C. Rovelli,
Loop space representation of quantum general relativity,
Nucl. Phys. B {\bf 331} (1990) 80.

\bibitem{nasiri2006psq}
S. Nasiri, Y. Sobouti and F. Taati,
Phase space quantum mechanics-Direct,
Journal of Mathematical Physics {\bf 47} (2006) 92106.

\bibitem{thiemann1996atg}
T. Thiemann,
An Account of transforms on A/G,
Acta Cosmologica {\bf 21} (1996) 145.

\bibitem{RevModPhys.52.515}
S. Weinberg,
Conceptual foundations of the unified theory of weak and
 electromagnetic interactions,
Rev. Mod. Phys. {\bf 52} (1980) 515.

\bibitem{RevModPhys.52.525}
A. Salam,
Gauge unification of fundamental forces,
Rev. Mod. Phys. {\bf 52} (1980) 525.

\bibitem{RevModPhys.52.539}
S.L. Glashow,
Towards a unified theory: Threads in a tapestry,
Rev. Mod. Phys. {\bf 52} (1980) 539.

\bibitem{cohen2006vsr}
A. Cohen and S. Glashow,
Very Special Relativity,
Physical Review Letters {\bf 97} (2006) 21601.

\bibitem{DiracPMQM}
P.A.M. Dirac,
The principles of quantum mechanics,
(The Clarendon Press, Oxford, 1930).

\bibitem{RevModPhys.20.367}
R.P. Feynman,
Space-Time Approach to Non-Relativistic Quantum Mechanics,
Rev. Mod. Phys. {\bf 20} (1948) 367.

\bibitem{seidewitz2007fsp}
E. Seidewitz,
Foundations of a spacetime path formalism for relativistic quantum mechanics,
Journal of Mathematical Physics {\bf 47} (2006) 112302.

\bibitem{nambu1950upt}
Y. Nambu,
The Use of the Proper Time in Quantum Electrodynamics I,
Progress of Theoretical Physics {\bf 5} (1950) 82.

\end{thebibliography}

\end{document}